\documentclass[reprint,aps,prl,letter,fleqn]{revtex4-2}
\usepackage{graphicx,xcolor,hyperref}
\graphicspath{{./figs/}}
\usepackage{bm}
\usepackage{amsmath,amsfonts,amssymb,braket}
\usepackage[normalem]{ulem}
\renewcommand{\epsilon}{\varepsilon}

\hypersetup{colorlinks=true,breaklinks,linkcolor=blue,urlcolor=blue,citecolor=blue}

\usepackage[version=4]{mhchem}
\newcommand{\cecuge}{\ce{Ce2CuGe6}}

\begin{document}

\title{Nonlocal Kondo-exchange-driven intrinsic anomalous Hall effect \\
in localized-$4f$ antiferromagnetic metals}

\author{Akimitsu Kirikoshi}
\author{Junya Otsuki}
\affiliation{
 Research Institute for Interdisciplinary Science, Okayama University, Okayama 700-8530, Japan
}

\date{\today}

\begin{abstract}
The anomalous Hall effect in antiferromagnetic metals has attracted considerable attention. Most known realizations involve itinerant $d$ electrons that simultaneously mediate charge transport and magnetic order. Here, we focus on $f$-electron materials, where localized magnetic moments and conduction electrons are hosted in different orbitals. We develop a theoretical framework to describe the impact of localized antiferromagnetic order on itinerant electrons. Applying this approach to the recently discovered $4f$ antiferromagnetic metal {\cecuge}, we identify the origin of both the intrinsic anomalous Hall conductivity and the spin splitting of the energy bands as spin-dependent intersite hopping induced by nonlocal Kondo exchange coupling, rather than a Zeeman-type effective field acting locally on the conduction electrons.
\end{abstract}

\maketitle

\textit{Introduction}---The anomalous Hall effect (AHE), long studied in ferromagnetic metals~\cite{nagaosaAnomalousHallEffect2010}, is now actively explored in antiferromagnetic (AFM) metals with little or no net magnetization~\cite{nakatsujiTopologicalMagnetsTheir2022,smejkalAnomalousHallAntiferromagnets2022}.
Representative examples include
\ce{Mn3Sn}~\cite{nakatsujiLargeAnomalousHall2015},
\ce{RuO2}~\cite{smejkalCrystalTimereversalSymmetry2020,Feng2022},
$\alpha$-Mn~\cite{akiba2020},
\ce{MnTe}~\cite{Betancourt2023},
and NbMnP~\cite{kotegawaLargeAnomalousHall2023,kotegawaLargeSpontaneousHall2024,araiIntrinsicAnomalousHall2024}.
Theoretically, the intrinsic contribution of the AHE is described by Berry curvature in momentum space~\cite{karplusHallEffectFerromagnetics1954,sundaramWavepacketDynamicsSlowly1999,haldaneBerryCurvatureFermi2004,ohgataIntrinsicAnomalousHall2025}.
In AFM metals, Berry curvature emerges from the interplay of crystal symmetry, magnetic sublattices, and spin-orbit coupling~\cite{Solovyev1997,Tomizawa2009,chenAnomalousHallEffect2014,suzukiClusterMultipoleTheory2017,naka2020,naka2022,attiasIntrinsicAnomalousHall2024,golubinskiiAnomalousHallEffect2026}, and is often accompanied by anisotropic momentum-dependent spin splitting of energy bands~\cite{Ahn2019,Naka2019,HayamiSpinSplit2019,smejkalCrystalTimereversalSymmetry2020}.

Most microscopic theories of the intrinsic AHE have been developed for itinerant AFM metals based on the Stoner picture.
A qualitatively different situation arises in localized-moment metals.
{\cecuge} is one such material that exhibits a large anomalous Hall conductivity (AHC) in the collinear AFM phase with localized $4f$ moments~\cite{kotegawaLargeAnomalousHall2024}.
Unlike itinerant AFM metals, the ordered moments reside on localized $4f$ electrons, whereas charge transport is carried mainly by conduction electrons derived from ligand orbitals such as \ce{Ge}~$4p$.
The separation between magnetism and transport carriers raises a fundamental question: how do localized magnetic moments generate a substantial Berry curvature in conduction-electron bands?
A natural expectation is that the ordered moments act on the conduction electrons as a molecular field, producing a Zeeman-like spin splitting.
We shall reveal, however, that such a naive picture is insufficient to account for the large AHC, calling for a microscopic description beyond the conventional Stoner picture.

\begin{figure}[bt]
  \centering
  \includegraphics[width=\linewidth]{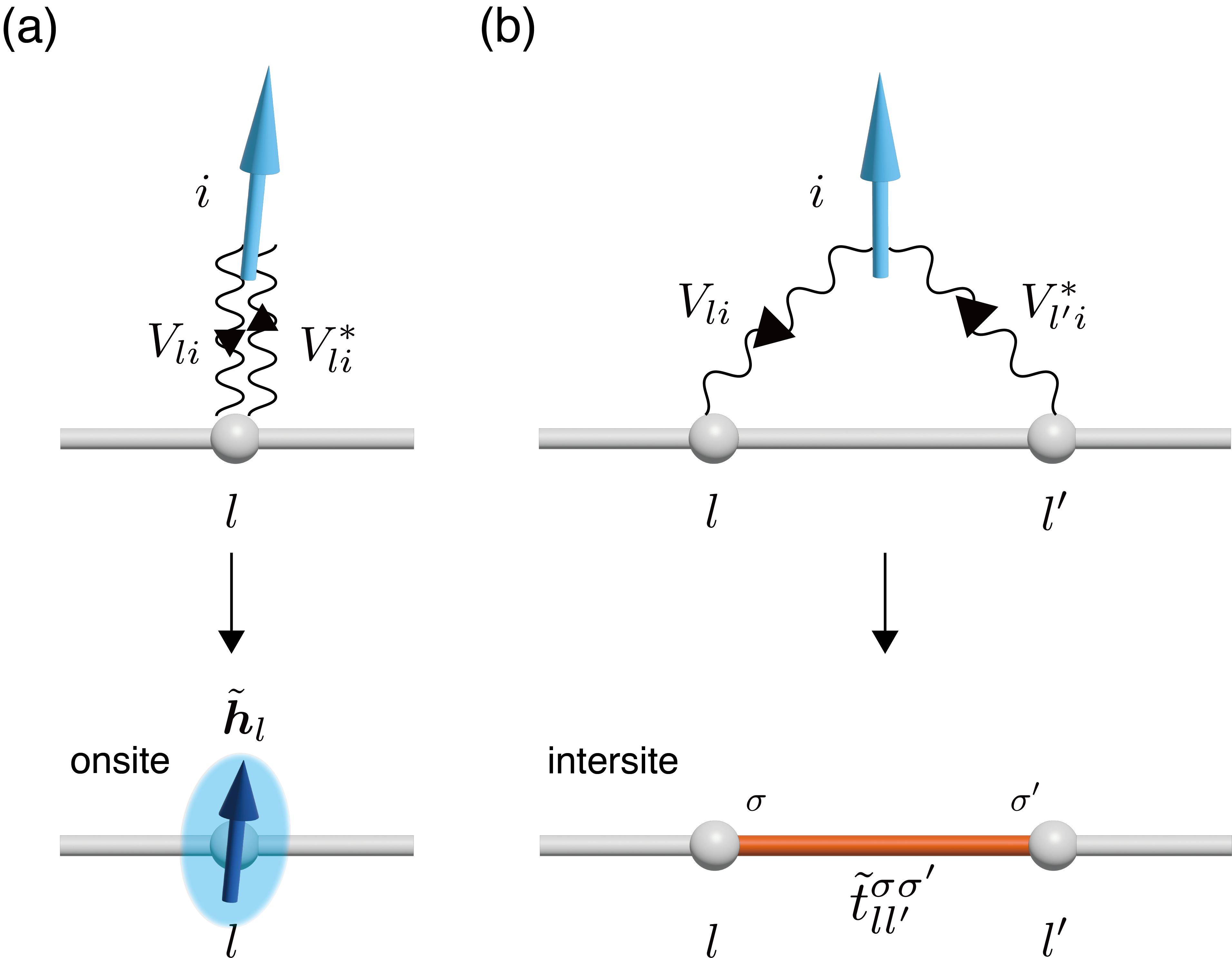}
  \caption{Schematic illustration of virtual $c$--$f$ hybridization processes and their consequences in the localized-$4f$ regime:
  (a) the local effective field $\tilde{\bm{h}}_{l}$, and (b) the nonlocal spin-dependent hopping $\tilde{t}_{ll^{\prime}}^{\sigma\sigma^{\prime}}$.
  Upper panels show representative processes, and lower panels show the corresponding effective terms. 
  The blue arrows and white spheres denote the $4f$-electron spins and the conduction-electron sites, respectively.
  }
  \label{fig:virtual_process}
\end{figure}

The effect of magnetically ordered localized moments on conduction electrons is described as follows.
One well-known consequence is the polarization of conduction-electron spins, giving rise to Friedel oscillations and the Ruderman-Kittel-Kasuya-Yosida (RKKY) interaction. 
This effect arises from the virtual hybridization process illustrated in Fig.~\ref{fig:virtual_process}(a), in which a conduction electron returns to its original site via the $f$ orbital.
Another process, shown in Fig.~\ref{fig:virtual_process}(b), mediates an effective hopping between different conduction-electron sites. 
Importantly, because the intermediate $4f$ state is subject to strong spin-orbit coupling, the hopping amplitude can be spin-dependent and may even involve a spin-flip process.
Through an application to {\cecuge}, we demonstrate that the latter process, i.e., the spin-dependent effective hopping, is the origin of the AHC in localized-$4f$ antiferromagnets.

\textit{Effective Hamiltonian}---Our starting point is a tight-binding Hamiltonian consisting of conduction and $4f$ electrons.
The Hamiltonian matrix for each momentum $\bm{k}$ is given by
\begin{equation}
  \hat{H}_{\mathrm{TB}}(\bm{k})=
  \begin{pmatrix}
    \hat{H}_{c}(\bm{k}) & \hat{V}(\bm{k})
    \\
    \hat{V}^{\dagger}(\bm{k}) & \hat{H}_{f}(\bm{k})
  \end{pmatrix},
  \label{eq:decomposition_hamiltonian}
\end{equation}
where the hat denotes the matrix in internal degrees of freedom, such as spin and orbital.
$\hat{H}_{c}(\bm{k})$ and $\hat{H}_{f}(\bm{k})$ describe the matrix elements within conduction electrons and $4f$ electrons, respectively, and $\hat{V}(\bm{k})$ is the $c$--$f$ hybridization.

We consider correlations among $f$ electrons.
In the strong-coupling regime where $f$ electrons are well localized, the energy bands around the Fermi level are predominantly of conduction-electron character.
The influence of the $f$ electrons can be incorporated by the effective Hamiltonian for conduction electrons given by
\begin{equation}
  \hat{H}_{c}^{\mathrm{eff}}(\bm{k})=\hat{H}_{c}(\bm{k})+\hat{V}(\bm{k})\hat{g}^{\mathrm{H}}_{f}(\bm{k},0)\hat{V}^{\dagger}(\bm{k}),
  \label{eq:effective_hamiltonian_c}
\end{equation}
where $\hat{g}_{f}(\bm{k},\omega)$ is defined by  
\begin{equation}
  \hat{g}_{f}(\bm{k},\omega)=[(\omega+\mu)\hat{1}-\hat{H}_{f}(\bm{k})-\hat{\Sigma}_{\mathrm{loc}}(\omega)]^{-1}.
  \label{eq:resolvent_f}
\end{equation}
Here, $\mu$ denotes the chemical potential, $\hat{1}$ is a unit matrix, and $\hat{\Sigma}_{\mathrm{loc}}(\omega)$ is the self-energy of $f$ electrons.
The $\bm{k}$-dependence of $\hat{\Sigma}_{\mathrm{loc}}(\omega)$ is neglected because the energy dependence plays the major role in the strong-coupling regime~\cite{georgesDynamicalMeanfieldTheory1996}.
The superscript H indicates that the Hermitian part of the self-energy is taken as
$\hat{\Sigma}^{\mathrm{H}}_{\mathrm{loc}}(\omega)=[\hat{\Sigma}_{\mathrm{loc}}(\omega)+\hat{\Sigma}_{\mathrm{loc}}^{\dag}(\omega)]/2$
to ensure the Hermiticity of $\hat{H}_{c}^{\mathrm{eff}}(\bm{k})$.
The derivation of Eq.~\eqref{eq:effective_hamiltonian_c} from the conduction electron Green's function is provided in Supplemental Material~\cite{SM}(Refs.~\cite{georgesDynamicalMeanfieldTheory1996,kotliarElectronicStructureCalculations2006,schriefferRelationAndersonKondo1966,coqblinExchangeInteractionAlloys1969,matsudaLargeSpontaneousHall2026,koepernikFullpotentialNonorthogonalLocalorbital1999,eschrigTightbindingModelsIronbased2009,PhysRevB.107.235135,anisimovFirstprinciplesCalculationsElectronic1997,shinaokaDCoreIntegratedDMFT2021,parcolletTRIQSToolboxResearch2015,aichhornTRIQSDFTToolsTRIQS2016,andrey_e_antipov_2017_825870,herbst$4f$ExcitationEnergies1976,herbstRelativisticCalculations$4f$1978,lochtStandardModelRare2016,tanakaTheoryAcAnomalous2008,parkerDiagrammaticApproachNonlinear2019,michishitaEffectsRenormalizationNonHermiticity2021,shinaokaCompressingGreensFunction2017,wallerbergerSparseirOptimalCompression2023,liSparseSamplingApproach2020,xiaoBerryPhaseEffects2010}).

The second term in Eq.~\eqref{eq:effective_hamiltonian_c} yields the correction that describes the effect of the localized $f$ moments.
Here, $\hat{g}_f(\bm{k},\omega)$ is approximated by its zero-energy value, and the spectral broadening is neglected.
These approximations are expected to be valid for describing electronic structure near the Fermi level.
Their accuracy will be assessed later by comparing the AHC evaluated with and without the effective model description.
We also note that our approach, which exploits the localized nature of $f$ electrons, is complementary to the itinerant treatment of $f$ electrons applied to AHE in ferromagnets~\cite{kohnoTheoryAnomalousHall1993,kontaniTheoryAnomalousHall1994}.

To elucidate the role of the correction term in Eq.~\eqref{eq:effective_hamiltonian_c}, we consider a simple case without orbital degrees of freedom.
Replacing $\hat{\Sigma}_{\mathrm{loc}}(\omega)$ with the analytic expression in the atomic limit, the effective Hamiltonian in Eq.~\eqref{eq:effective_hamiltonian_c} reduces to the Kondo lattice model with the localized spin replaced by its thermal average. Its second-quantized expression $\mathcal{H}_{c}^{\mathrm{eff}}$ is given by 
\begin{equation}
  \mathcal{H}_{c}^{\mathrm{eff}}\to\sum_{ll^{\prime}}\sum_{\sigma}t_{ll^{\prime}}c_{l\sigma}^{\dag}c_{l^{\prime}\sigma}+\sum_{ll^{\prime}}\sum_{i}J_{\mathrm{K},ll^{\prime}}^{i}\braket{\bm{S}_{i}}\cdot\bm{s}_{ll^{\prime}},
  \label{eq:generalized_Kondo_lattice_model}
\end{equation}
where $c_{l\sigma}^{\dag}$ and $c_{l\sigma}$ denote creation and annihilation operators for conduction electrons at site $l$ with spin $\sigma$, and the coefficient $t_{ll^{\prime}}$ is the spin-independent hopping amplitude from $\hat{H}_{c}$.
$\braket{\bm{S}_{i}}$ denotes the thermal average of the localized $f$-electron spin at site $i$.
The operator $\bm{s}_{ll^{\prime}}$ is defined by $\bm{s}_{ll^{\prime}}=\sum_{\sigma\sigma^{\prime}}(\bm{\tau}_{\sigma\sigma^{\prime}}/2)c_{l\sigma}^{\dag}c_{l^{\prime}\sigma^{\prime}}$, where $\hat{\bm{\tau}}=(\hat{\tau}_{x},\hat{\tau}_{y},\hat{\tau}_{z})$ is the Pauli matrices.
It is important that the Kondo exchange coupling $J_{\mathrm{K},ll^{\prime}}^{i}\propto V_{li}V_{l^{\prime}i}^{*}$ is bond-resolved, originating from nonlocal $c$--$f$ hybridization.

For $l=l^{\prime}$, the operator $\bm{s}_{ll}$ reduces to the local spin density, and the second term in Eq.~\eqref{eq:effective_hamiltonian_c} acts as a molecular field $\sum_{l} \tilde{\bm{h}}_{l}\cdot\bm{s}_{ll}$ on conduction electrons with $\tilde{\bm{h}}_{l} \equiv\sum_{i}J_{\mathrm{K},ll}^{i}\braket{\bm{S}_{i}}$.
On the other hand, the $l\neq l^{\prime}$ contribution provides a spin-dependent correction $\tilde{t}_{ll^{\prime}}^{\sigma\sigma^{\prime}} \equiv \sum_{i}J_{\mathrm{K},ll^{\prime}}^{i}\braket{\bm{S}_{i}}\cdot\bm{\tau}_{\sigma\sigma^{\prime}}/2$
to the bare hopping amplitude $t_{ll^{\prime}}$.
These local and nonlocal contributions are illustrated in the lower panel of Fig.~\ref{fig:virtual_process}.
The effective Hamiltonian in Eq.~\eqref{eq:effective_hamiltonian_c} thus provides a mean-field description of localized $f$ moments when $\hat{\Sigma}_{\mathrm{loc}}(\omega)$ is evaluated in the atomic limit.
Additional correlation effects can be incorporated through a more elaborate evaluation of $\hat{\Sigma}_{\mathrm{loc}}(\omega)$ using, e.g., dynamical mean-field theory~\cite{georgesDynamicalMeanfieldTheory1996,kotliarElectronicStructureCalculations2006}.

\textit{Intrinsic AHE in \cecuge}---We analyze {\cecuge}, a localized-$4f$ AFM metal exhibiting a large AHC~\cite{kotegawaLargeAnomalousHall2024}.
This compound undergoes an AFM transition below $T_{\mathrm{N}}\simeq 15$~K~\cite{luMagneticAnomaliesSpinGlassLike2011,qiCrystalMagneticStructures2019}.
The ordered moments form a collinear AFM structure along the $b$ axis~\cite{qiCrystalMagneticStructures2019}, corresponding to the magnetic point group $m^{\prime}m^{\prime}m$.
The mirror plane is perpendicular to the $c$ axis, thus the AHE in the $a$--$b$ plane is allowed~\cite{yatsushiroMultipoleClassification1222021}.
Although the experimentally observed AHC is dominated by extrinsic contributions, the intrinsic contribution is relatively large, with an estimated magnitude of $\sim80~\Omega^{-1}\mathrm{cm}^{-1}$ compared with a total of about 550~$\Omega^{-1}\mathrm{cm}^{-1}$~\cite{kotegawaLargeAnomalousHall2024}.
The Kondo temperature is estimated to be $T_{\mathrm{K}} \simeq 1$~K~\cite{tsengAntiferromagneticSpinWave2003}, which justifies a local-moment approach for {\cecuge}.

First, we perform nonmagnetic DFT calculations from the experimental crystal structure parameters~\cite{matsudaLargeSpontaneousHall2026} and construct the tight-binding model $\hat{H}_{\mathrm{TB}}$ using \texttt{FPLO}~\cite{koepernikFullpotentialNonorthogonalLocalorbital1999,eschrigTightbindingModelsIronbased2009,PhysRevB.107.235135}.
For simplicity, we neglect the $j=7/2$ states of $4f$ electrons, which lie 0.3\,eV above the $j=5/2$ states because of spin--orbit coupling.
The crystalline-electric-field (CEF) ground state is the $j_z=\pm1/2$ doublet, and the magnetic easy axis lies in the $a$--$b$ plane, consistent with the experiments~\cite{qiCrystalMagneticStructures2019}.
Next, we incorporate the strong correlation effects of the $f$ electrons through the rotationally invariant atomic interactions with the Coulomb repulsion $U=6.17$\,eV and the Hund's coupling $J_\textrm{H}=0.79$\,eV.
We compute the self-energy of the $f$ electrons in the Matsubara frequency domain $\hat{\Sigma}_{\mathrm{loc}}(z)$ within the Hubbard-I approximation to describe the completely localized limit.
At this stage, we introduce a molecular field for $4f$ electrons to reproduce the collinear AFM structure with moments along the $b$ axis.
To be consistent with experimental observations~\cite{qiCrystalMagneticStructures2019}, we adjust the CEF potential so that the magnetic moments along the $b$ axis do not induce the $a$-axis component.
We use the open-source software \texttt{DCore}~\cite{shinaokaDCoreIntegratedDMFT2021}, implemented with \texttt{TRIQS}~\cite{parcolletTRIQSToolboxResearch2015} and \texttt{DFTTools}~\cite{aichhornTRIQSDFTToolsTRIQS2016} libraries, together with the exact diagonalization program~\texttt{pomerol}~\cite{andrey_e_antipov_2017_825870}.
Further detailed calculation procedure is described in the Supplemental Material~\cite{SM}.

The results exhibit a fully polarized AFM state along the $b$ axis with a weak canting magnetization $m_z \approx 2.6\times 10^{-2}~\mu_{\mathrm{B}}/\ce{Ce}$ along the $c$ axis.
This result is consistent with the experimental value $3\times 10^{-2}~\mu_{\mathrm{B}}/\ce{Ce}$~\cite{kotegawaLargeAnomalousHall2024}.
The appearance of the canting magnetic moments is attributed to the anisotropic CEF potential of $4f$ electrons.

\begin{figure}[tb]
  \centering
  \includegraphics[width=\linewidth]{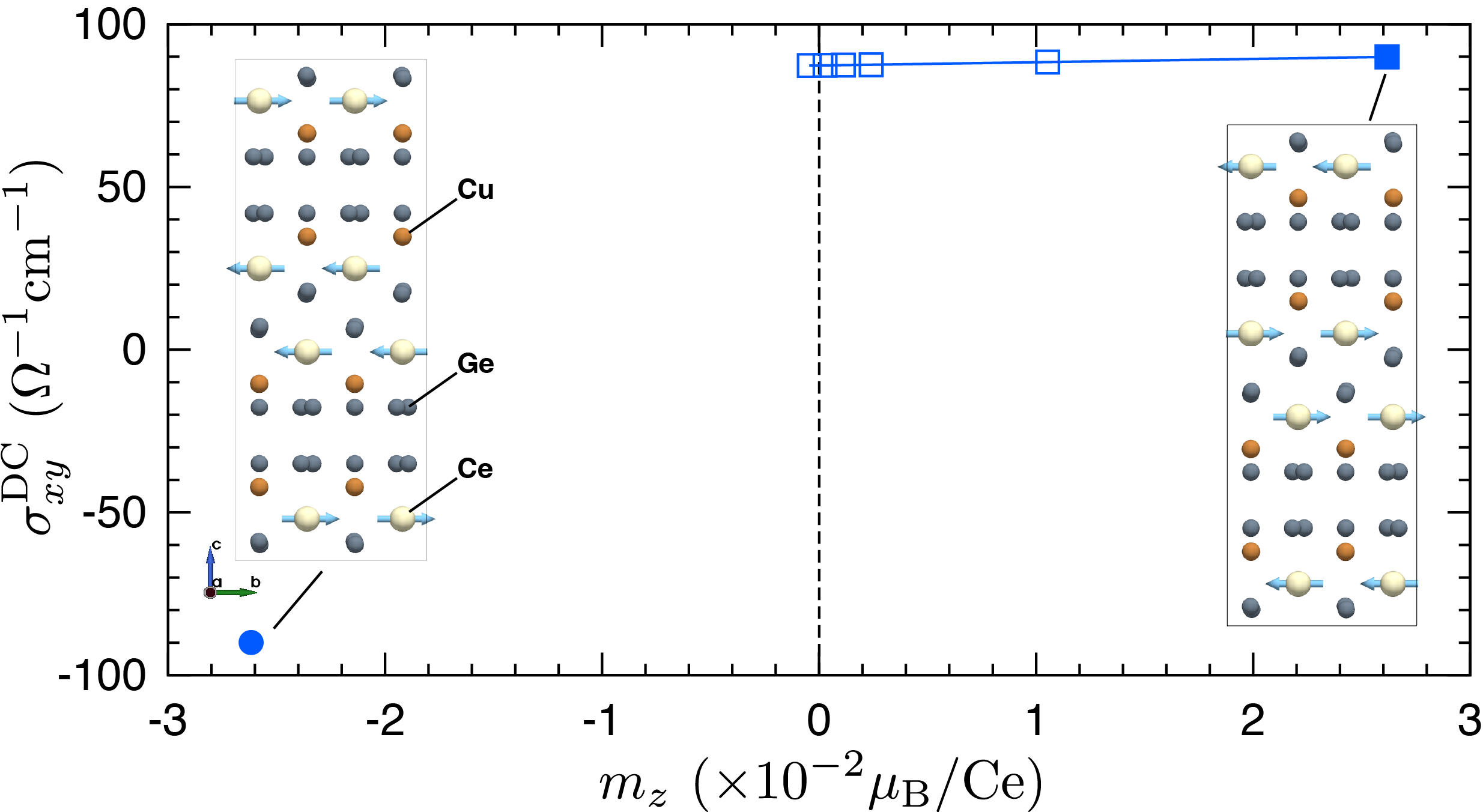}
  \caption{Intrinsic AHC $\sigma_{xy}^{\mathrm{DC}}$ versus canting magnetization $m_z$.
  Filled symbols represent the results with the molecular field for collinear antiferromagnetic (AFM) order along the $b$ axis, as shown in the insets.
  Open symbols denote canting tuned by a uniform field along the $c$ axis, with a fixed in-plane magnetic configuration.
  The crystal structures are generated by QtDraw~\cite{kusunoseSymmetryadaptedModelingMolecules2023}.
  }
  \label{fig:canting_AHC}
\end{figure}

Before applying the effective Hamiltonian, we evaluate the intrinsic AHC directly from the Green's function, including the self-energy. 
We employ the current--current correlation function within the bubble diagram~\cite{tanakaTheoryAcAnomalous2008,parkerDiagrammaticApproachNonlinear2019,michishitaEffectsRenormalizationNonHermiticity2021}.
Figure~\ref{fig:canting_AHC} shows the calculated intrinsic AHC $\sigma_{xy}^{\mathrm{DC}}$ plotted against the canting magnetization $m_z$.
Here, we set the $x$, $y$, and $z$ axes to the $a$, $b$, and $c$ axes, respectively.
We obtain $|\sigma_{xy}^{\mathrm{DC}}|\simeq 90~\Omega^{-1}\mathrm{cm}^{-1}$ (filled square), consistent with experiment~\cite{kotegawaLargeAnomalousHall2024}.
Reversing the in-plane magnetic-moment direction flips the sign of the AHC as well as $m_z$ (filled circle).
To disentangle the contribution of the collinear AFM structure from that of the canting magnetization $m_z$, we apply a uniform magnetic field along the $c$ axis and suppress $m_z$.
As shown by the open squares in Fig.~\ref{fig:canting_AHC}, the AHC remains nearly unchanged even when $m_z$ is reduced to zero.
These results reveal that the intrinsic AHE stems from the Berry curvature generated by the collinear AFM order, with negligible contribution from the canting-induced ferromagnetic component or scalar spin chirality~\cite{yeBerryPhaseTheory1999,ohgushiSpinAnisotropyQuantum2000,tataraChiralityDrivenAnomalousHall2002,brunoTopologicalHallEffect2004}.

\begin{figure}[tb]
  \centering
  \includegraphics[width=\linewidth]{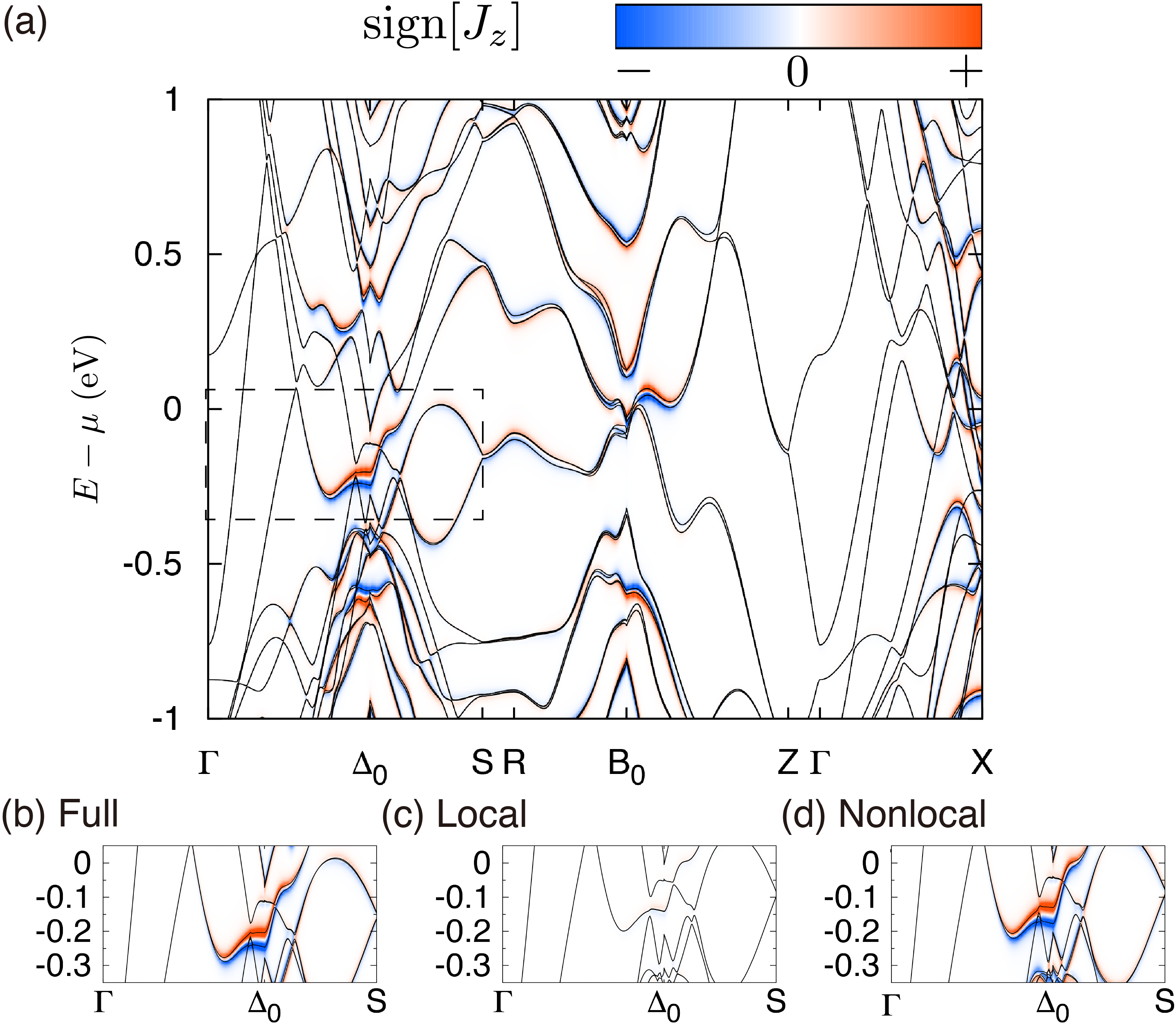}
  \caption{(a) Band structure from the effective Hamiltonian.
  The color map shows the atomic-scale total angular momentum polarization $J_{z}$ (sign indicated by the color bar at the top).
  (b) Zoom-in of the dashed region in (a).
  Panels (c) and (d) show the results obtained by retaining only the \ce{Ge}~$4p$ local effective field and \ce{Ge}~$4p$ nonlocal effective hopping, respectively.
  }
  \label{fig:band_structure_effective_hamiltonian}
\end{figure}

Now, we apply the effective-Hamiltonian approach in Eq.~\eqref{eq:effective_hamiltonian_c} and examine the microscopic mechanism underlying intrinsic AHC.
Figure~\ref{fig:band_structure_effective_hamiltonian}(a) shows the band structure of conduction electrons obtained from $\hat{H}_{c}^{\mathrm{eff}}(\bm{k})$.
The color map indicates the $z$ component of the total angular momentum, $J_z$.
We observe the spin splitting of the band, whose magnitude is about 40\,meV at maximum near the Fermi level around the $X$ and $\Delta_0$ points.
Since the first term in Eq.~\eqref{eq:effective_hamiltonian_c} is constructed for the paramagnetic state, the spin polarization of the conduction electrons is induced by the $c$--$f$ hybridization.

\begin{figure}[tb]
  \centering
  \includegraphics[width=\linewidth]{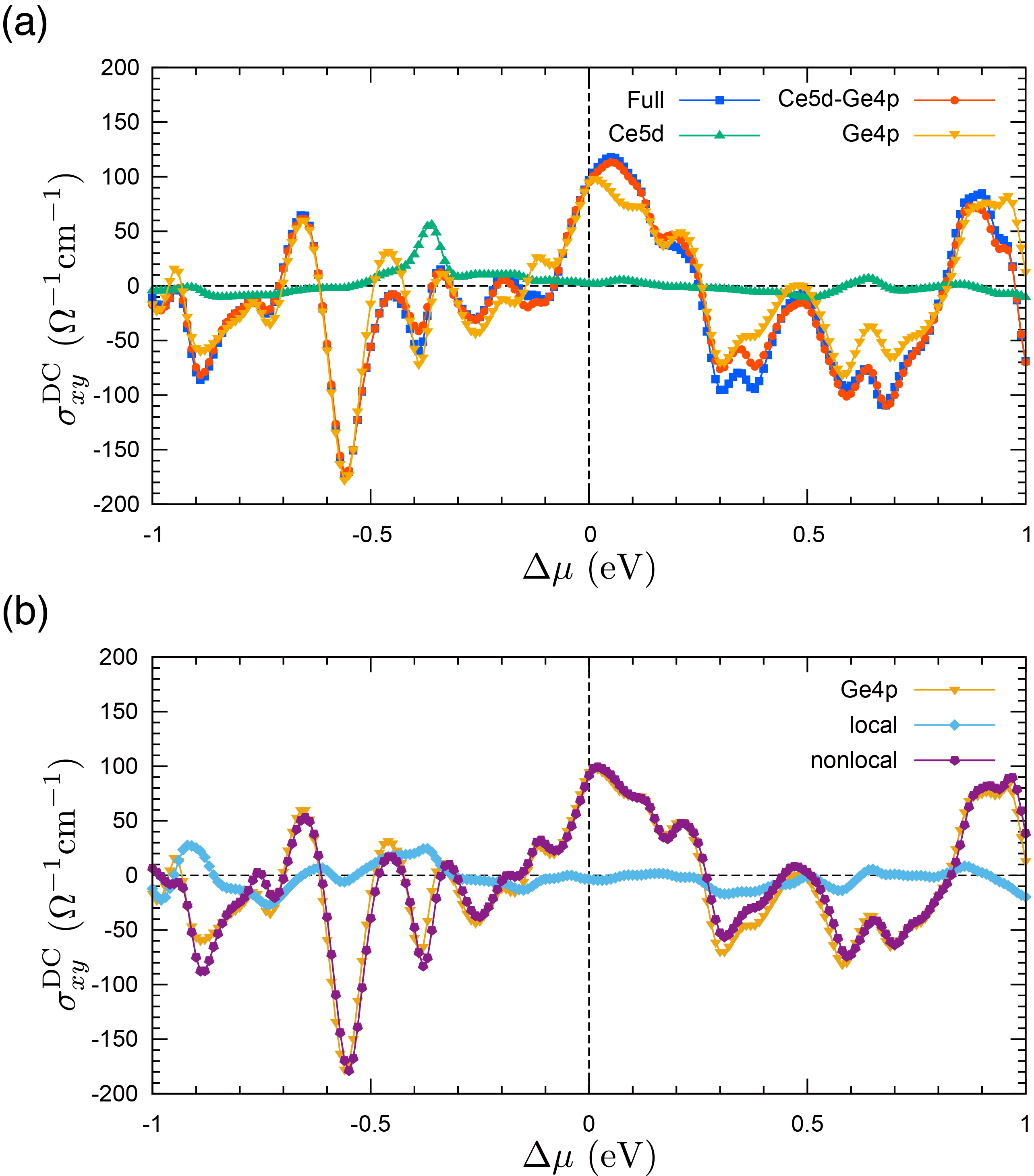}
  \caption{Intrinsic AHC $\sigma_{xy}^\textrm{DC}$ versus a chemical potential shift $\Delta \mu$ from the Fermi level, with decomposition into several contributions. 
  (a) A decomposition into \ce{Ce}~$5d$ and \ce{Ge}~$4p$ contributions. The blue squares labeled with `Full' correspond to the result of the effective Hamiltonian.
  (b) A decomposition of the \ce{Ge}~$4p$ contribution into local effective-field and nonlocal effective-hopping components.}
  \label{fig:LC_band_coupling}
\end{figure}

We calculate the intrinsic AHC from the effective tight-binding band using linear response theory as follows:
\begin{equation}
  \sigma_{xy}^{\mathrm{DC}}=-\frac{e^{2}}{\hbar}\int\frac{d\bm{k}}{(2\pi)^{3}}\sum_{n}f_{n\bm{k}}\Omega_{n\bm{k}}^{z},
  \label{eq:AHC_effective_hamiltonian}
\end{equation}
where $f_{n\bm{k}}$ and $\Omega_{n\bm{k}}^{z}$ are the Fermi-Dirac distribution function and band-resolved Berry curvature for the $n$-th band at wavevector $\bm{k}$ in the Brillouin zone, respectively.
We obtain $\sigma_{xy}^{\mathrm{DC}}\simeq 97~\Omega^{-1}\mathrm{cm}^{-1}$, which reproduces the result from the full Green's-function calculation in Fig.~\ref{fig:canting_AHC}.
Figure~\ref{fig:LC_band_coupling}(a) shows $\sigma_{xy}^{\mathrm{DC}}$ as a function of the chemical-potential shift $\Delta\mu$ (labeled with `Full'). 
The peak position of the intrinsic AHC coincides with the spin polarization of the conduction electrons in Fig.~\ref{fig:band_structure_effective_hamiltonian}(a), which implies that the spin-split band near the Fermi energy is responsible for the intrinsic AHC.
Indeed, we confirmed that hot spots of the Berry curvature appear in the regions of large spin polarization near the $\Delta_{0}$ and $\mathrm{X}$ points in the Brillouin zone~\cite{SM}.

We now exploit the effective Hamiltonian to identify the orbitals responsible for the intrinsic AHC.
To this end, we selectively eliminate some matrix elements in the hybridization matrix $\hat{V}$ and isolate the contributions of specific orbitals.
Figure~\ref{fig:LC_band_coupling}(a) compares the AHC obtained by retaining only the \ce{Ce}~$5d$ orbitals, only the \ce{Ge}~$4p$ orbitals, or both, with that from the full-orbital calculation.
We find that the \ce{Ge}~$4p$ orbitals provide the dominant contribution to the AHC and \ce{Ce}~$5d$ orbitals yield only minor corrections away from the Fermi level.

We further resolve the \ce{Ge}~$4p$ contribution into the local effective field and the nonlocal effective hopping, corresponding to Figs.~\ref{fig:virtual_process}(a) and \ref{fig:virtual_process}(b), respectively.
This is implemented by retaining either the site-diagonal ($l=l'$) or off-diagonal ($l\neq l'$) components in the second term of Eq.~\eqref{eq:effective_hamiltonian_c}.
Figure~\ref{fig:LC_band_coupling}(b) shows the intrinsic AHC recalculated from the resulting effective Hamiltonians.
Remarkably, the results obtained from the nonlocal components account for almost the entire \ce{Ge}~$4p$ contributions. 
Therefore, we conclude that the microscopic origin of the large AHC is nonlocal spin-dependent hopping generated by $c$--$f$ hybridization, with only a minor contribution from the local effective field.
We can also identify the \ce{Ge}--\ce{Ce}--\ce{Ge} hopping path responsible for the intrinsic AHC~\cite{SM}.

Finally, we turn our attention back to the spin splitting of the energy bands. 
Figure~\ref{fig:band_structure_effective_hamiltonian}(b) focuses on the region enclosed by the dashed line in Fig.~\ref{fig:band_structure_effective_hamiltonian}(a), and Figs.~\ref{fig:band_structure_effective_hamiltonian}(c) and \ref{fig:band_structure_effective_hamiltonian}(d) show the corresponding results computed by retaining only the local and nonlocal contributions of \ce{Ge}~$4p$ orbitals.
Consistent with the AHC, the spin splitting is driven by the nonlocal contributions from \ce{Ge}~$4p$ bands.

\textit{Conclusion}---We developed an effective-model approach that incorporates the influence of localized magnetic order on itinerant charge carriers.
Applying it to the localized-$4f$ collinear AFM metal {\cecuge}, we showed that the intrinsic AHC is governed not by a local effective field or canting-induced ferromagnetism, but by spin-dependent intersite hopping generated by $c$--$f$ hybridization, reflecting the strong spin-orbit coupling of the $4f$ electrons.
This mechanism demonstrates that conduction bands can acquire Berry curvature through nonlocal Kondo exchange interactions, even in the absence of coherent $f$-electron quasiparticles. 
Our results establish nonlocal Kondo exchange as an essential ingredient for describing the AHE in localized-$4f$ antiferromagnets.

The present approach offers a simple and versatile scheme for analyzing transport phenomena beyond the AHE in materials with localized magnetism.
Promising future applications include nonlinear transverse transport in \ce{HoAgGe}~\cite{Miyamoto2025} and nonreciprocal transport in \ce{NdRu2Al10}~\cite{Sudo2026}, where symmetry considerations account for the existence of the response, but a quantitative microscopic understanding remains an important open challenge.

\acknowledgements{
  We thank H. Kotegawa, R. Hibino, and M. Naka for fruitful discussions.
  This work was supported by JSPS KAKENHI Grants No.~23H04869, No.~25K22013, and No.~26K00659.
  The computation in this work was performed using the facilities of the Supercomputer Center at the Institute for Solid State Physics, the University of Tokyo (Grants No.~2025-Ca-0086, No.~2025-Cb-0049, and 2026-Ca-0142).
}

\bibliography{manuscript_arXiv.bbl}

\newpage

\begin{widetext}
  \setcounter{page}{1}
  \setcounter{figure}{0}
  \setcounter{table}{0}
  \setcounter{equation}{0}

  \renewcommand{\thefigure}{S\arabic{figure}}
  \renewcommand{\thetable}{S\arabic{table}}

  \renewcommand{\theHfigure}{S\arabic{figure}}
  \renewcommand{\theHtable}{S\arabic{table}}
  
  \renewcommand{\theequation}{S\arabic{equation}}
  \renewcommand{\theHequation}{S\arabic{equation}}

\begin{center}
  \large\bf
  Supplemental Material for \\
  ``Nonlocal Kondo-exchange-driven intrinsic anomalous Hall effect \\
  in localized-$4f$ antiferromagnetic metals''
\end{center}

\section{Effective Hamiltonian for the conduction electrons}

\subsection{Derivation of the effective Hamiltonian}

We derive the effective Hamiltonian for the conduction electrons from the correlated-electron model Hamiltonian.
We treat the strong correlation effects of the $f$ electrons arising from the on-site Coulomb interaction using dynamical mean-field theory (DMFT)~\cite{georgesDynamicalMeanfieldTheory1996,kotliarElectronicStructureCalculations2006}.
The DMFT calculations yield the $f$-electron self-energy $\hat{\Sigma}_{\mathrm{loc}}(z)$.
Then, the single-particle Green's function is given by
\begin{equation}
  \hat{G}(\bm{k},z)=[(z+\mu)\hat{1}-\hat{H}_{\mathrm{TB}}(\bm{k})-\hat{\Sigma}_{\mathrm{loc}}(z)]^{-1}.
  \label{eq:Green_function_full_SM}
\end{equation}
Here, $z$ is the Matsubara frequency, $\mu$ is the chemical potential, $\hat{1}$ is the unit matrix, $\hat{H}_{\mathrm{TB}}(\bm{k})$ is the tight-binding Hamiltonian, and $\hat{\Sigma}_{\mathrm{loc}}(z)$ is the self-energy of the $f$ electrons.
We write $\hat{H}_{\mathrm{TB}}(\bm{k})$ in block form for the conduction electrons, the $f$ electrons, and the $c$--$f$ hybridization:
\begin{equation}
  \hat{H}_{\mathrm{TB}}(\bm{k})=
  \begin{pmatrix}
    \hat{H}_{c}(\bm{k}) & \hat{V}(\bm{k})
    \\
    \hat{V}^{\dagger}(\bm{k}) & \hat{H}_{f}(\bm{k})
  \end{pmatrix},
  \label{eq:decomposition_hamiltonian_SM}
\end{equation}
where $\hat{H}_{c}(\bm{k})$, $\hat{H}_{f}(\bm{k})$, and $\hat{V}(\bm{k})$ are the Hamiltonians for conduction and $f$ electrons, and the $c$--$f$ hybridization term, respectively.

Using the decomposition in Eq.~\eqref{eq:decomposition_hamiltonian_SM}, the single-particle Green's function in Eq.~\eqref{eq:Green_function_full_SM} can be written as
\begin{equation}
  \hat{G}(\bm{k},z)=
  \begin{bmatrix}
    (z+\mu)-\hat{H}_{c}(\bm{k}) & \hat{V}(\bm{k})
    \\
    \hat{V}^{\dagger}(\bm{k}) & (z+\mu)-\hat{H}_{f}(\bm{k})-\hat{\Sigma}_{\mathrm{loc}}(z)
  \end{bmatrix}^{-1}
  \equiv
  \begin{bmatrix}
    \hat{G}_{c}(\bm{k},z) & \hat{G}_{cf}(\bm{k},z)
    \\
    \hat{G}_{fc}(\bm{k},z) & \hat{G}_{f}(\bm{k},z)
  \end{bmatrix}.
\end{equation}
The single-particle Green's function of the conduction electrons $\hat{G}_{c}(\bm{k},z)$ is given by
\begin{equation}
  \hat{G}_{c}(\bm{k},z)=[(z+\mu)\hat{1}-\hat{H}_{c}(\bm{k})-\hat{V}(\bm{k})\hat{g}_{f}(\bm{k},z)\hat{V}^{\dagger}(\bm{k})]^{-1}
  \label{eq:Green_function_c_SM}
\end{equation}
with $\hat{g}_{f}(\bm{k},z)=[(z+\mu)\hat{1}-\hat{H}_{f}(\bm{k})-\hat{\Sigma}_{\mathrm{loc}}(z)]^{-1}$.
The last term in Eq.~\eqref{eq:Green_function_c_SM} can be interpreted as a self-energy term generated by the $c$--$f$ hybridization.
By replacing $z\to 0$ in Eq.~\eqref{eq:Green_function_c_SM}, which corresponds to the static mean-field approximation, we obtain the effective Hamiltonian for the conduction electrons in the main text.

\subsection{Relationship between the effective Hamiltonian and the Kondo lattice model}

We show that the effective Hamiltonian in the main text reduces to the Kondo lattice model with the localized spin replaced by its thermal average in the atomic limit.
If $f$ electrons are completely localized, in which case $\hat{H}_{f}(\bm{k})$ is independent of the wavenumber $\bm{k}$, $\hat{g}_{f}(\bm{k},z)$ reduces to the local Green's function within the Hubbard-I solution.
To simplify the discussion, we neglect the orbital degeneracy and sublattice degrees of freedom.
We also assume that there is one $f$ electron per site and that it is completely polarized along the $\bm{m}$ direction.
Then, the local Green's function of the $f$ electrons is given by
\begin{equation}
  \hat{g}_{f}(z)
  =\frac{1}{2}\left(\frac{1}{z+\Delta_{-}}+\frac{1}{z-\Delta_{+}}\right)
  +\frac{1}{2}\left(\frac{1}{z+\Delta_{-}}-\frac{1}{z-\Delta_{+}}\right)\bm{m}\cdot\hat{\bm{\tau}}
\end{equation}
where $\Delta_{-}=\mu-\epsilon_{f}$ and $\Delta_{+}=-\mu+\epsilon_{f}+U$ ($\epsilon_{f}$ is the energy level of the $4f$ orbitals).
Substituting $\hat{g}_{f}(z\to 0)$ into Eq.~(2) in the main text, we obtain the effective Hamiltonian for the conduction electrons as
\begin{equation}
  \mathcal{H}_{c}^{\mathrm{eff}}=\sum_{ll^{\prime}}\sum_{\sigma}\left[t_{ll^{\prime}}+\sum_{i}\frac{1}{2}V_{li}V_{l^{\prime}i}^{*}\left(\frac{1}{\Delta_{-}}-\frac{1}{\Delta_{+}}\right)\right]c_{l\sigma}^{\dag}c_{l^{\prime}\sigma}
  +\sum_{ll^{\prime}}\sum_{\sigma\sigma^{\prime}}\sum_{i}2V_{li}V_{l^{\prime}i}^{*}\left(\frac{1}{\Delta_{-}}+\frac{1}{\Delta_{+}}\right)\frac{\bm{m}}{2}\cdot\frac{\bm{\tau}_{\sigma\sigma^{\prime}}}{2}c_{l\sigma}^{\dag}c_{l^{\prime}\sigma^{\prime}}.
  \label{eq:effective_hamiltonian_c_SM}
\end{equation}
in the second-quantized representation.
$t_{ll^{\prime}}$ is the hopping amplitude between sites $l$ and $l^{\prime}$ from $\hat{H}_{c}$.
The second term in the square brackets is the potential-scattering term due to the $c$--$f$ hybridization, which we omitted in the main text for simplicity.
The last term is the Kondo exchange coupling arising from the $c$--$f$ hybridization.
By defining the Kondo exchange coupling as
\begin{equation}
  J_{ll^{\prime}}^{i}=2V_{li}V_{l^{\prime}i}^{*}\left(\frac{1}{\Delta_{-}}+\frac{1}{\Delta_{+}}\right),
\end{equation}
$\braket{\bm{S}_{i}}=\bm{m}/2$, and $\bm{s}_{ll^{\prime}}=\sum_{\sigma\sigma^{\prime}}c_{l\sigma}^{\dag}(\bm{\tau}_{\sigma\sigma^{\prime}}/2)c_{l^{\prime}\sigma^{\prime}}$, we obtain the generalized Kondo lattice model with thermally averaged localized spins.
This model is easily extended to cases where the $c$--$f$ hybridization or magnetic order depends on internal degrees of freedom, such as sublattice or orbital.

The downfolding form $\hat{V}\hat{g}_{f}(z\!\to\!0)\hat{V}^{\dagger}$ is consistent with the second-order low-energy reduction of the Schrieffer--Wolff transformation~\cite{schriefferRelationAndersonKondo1966}.
Therefore, the present model corresponds to the Schrieffer--Wolff / Coqblin--Schrieffer exchange-interaction framework in a thermally averaged local-moment background.
Accordingly, the exchange-interaction term in Eq.~\eqref{eq:effective_hamiltonian_c_SM} is the static counterpart of the Kondo-type exchange interaction obtained by eliminating high-energy $f$ charge excitations~\cite{coqblinExchangeInteractionAlloys1969}.

\section{DFT with strong correlation effects of $f$ electrons}

In this section, we describe the DFT+Hubbard-I setup for \cecuge.
The space group of the crystal is $Cmce$.
Figure~\ref{fig:electronic_struct_SM}(a) shows the crystal and dominant magnetic structures of \cecuge.
In the following, we set the $x$, $y$, and $z$ axes to the $a$, $b$, and $c$ axes, respectively.
The structure parameters are as follows:
The lattice constants are $a=8.40700~\mathrm{Å}$, $b=8.12810~\mathrm{Å}$, and $c=21.53050~\mathrm{Å}$.
The Wyckoff positions of the atoms in the unit cell of \cecuge~are summarized in Table~\ref{tab:wyckoff_positions_SM}~\cite{matsudaLargeSpontaneousHall2026}.

\begin{table}[tb]
  \centering
  \caption{Wyckoff positions of the atoms in the unit cell of \cecuge.}
  \begin{tabular}{c|c|c|c|c|c}
    \hline
    Atom & Wyckoff position & $x$ & $y$ & $z$ & site symmetry\\
    \hline
    \ce{Ce} & $16g$ & $0.25076$ & $0.37488$ & $0.08328$ & $1$ \\
    \ce{Cu} & $8f$ & $0$ & $0.12461$ & $0.14687$ & $m..$ \\
    \ce{Ge} & $8f$ & $0$ & $0.12829$ & $0.03221$ & $m..$ \\
    \ce{Ge} & $8f$ & $0$ & $0.12231$ & $0.46160$ & $m..$ \\
    \ce{Ge} & $8f$ & $0$ & $0.40565$ & $0.19356$ & $m..$ \\
    \ce{Ge} & $8f$ & $0$ & $0.34426$ & $0.30623$ & $m..$ \\
    \ce{Ge} & $16g$ & $0.27773$ & $0.12525$ & $0.19387$ & $1$ \\
    \hline
  \end{tabular}
  \label{tab:wyckoff_positions_SM}
\end{table}

\subsection{Electronic band structure}

\begin{figure}[tb]
  \centering
  \includegraphics[width=\linewidth]{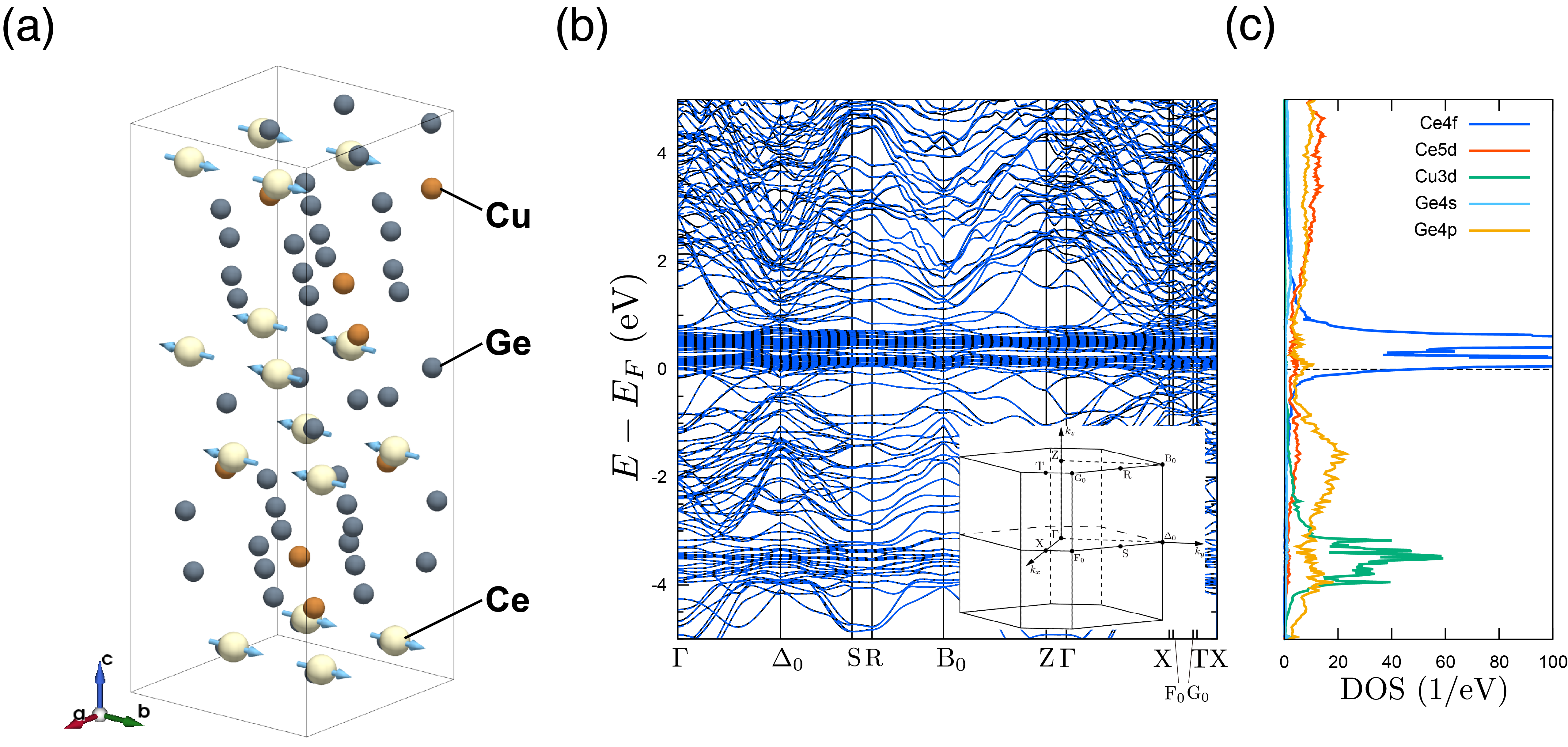}
  \caption{(a) Crystal structure of \cecuge.
  Arrows on the \ce{Ce} sites indicate the dominant magnetic order.
  (b) Band structure in the paramagnetic state.
  Black solid lines show DFT bands and blue dashed lines show the tight-binding bands.
  The inset shows the Brillouin zone.
  (c) Density of states for the orbitals included in the tight-binding model.}
  \label{fig:electronic_struct_SM}
 \end{figure}

We perform DFT calculations under the paramagnetic state using the full-potential local-orbital package FPLO~\cite{koepernikFullpotentialNonorthogonalLocalorbital1999} with the generalized-gradient approximation (GGA) and full-relativistic treatment of the valence electrons.
The charge density is converged self-consistently on a $14\times 14\times 6$ $\bm{k}$-point mesh.
Figure~\ref{fig:electronic_struct_SM}(b) shows the band structure and the density of states for the paramagnetic state.
The $4f$ bands are located near the Fermi energy in the absence of correlations.
We construct a tight-binding Hamiltonian from projector Wannier functions within FPLO~\cite{eschrigTightbindingModelsIronbased2009,PhysRevB.107.235135} to reproduce the DFT dispersion near the Fermi energy.
We set the energy window from $-13$ to $6$~eV to construct the tight-binding Hamiltonian.
The resulting 424-orbital model (including spin degrees of freedom) comprises \ce{Ce}~$4f$, \ce{Ce}~$5d$, \ce{Cu}~$3d$, \ce{Ge}~$4s$, and \ce{Ge}~$4p$ orbitals, as shown in Fig.~\ref{fig:electronic_struct_SM}(c).
We discuss the crystalline electric field (CEF) states of $4f$ electrons.
The local site symmetry of \ce{Ce} is $1$, as shown in Table~\ref{tab:wyckoff_positions_SM}, so that the CEF Hamiltonian for the $4f$ $j=5/2$ state is given by
\begin{equation}
  \hat{H}_{\mathrm{CEF}}=\sum_{m=-2}^{2}B_{2m}O_{2m}+\sum_{m=-4}^{4}B_{4m}O_{4m}.
\end{equation}
where $B_{lm}$ ($l=2,4$ and $m=-l,-l+1,\ldots,l-1,l$) are the CEF parameters and $O_{lm}$ are the cubic harmonics.
Since all $B_{lm}$ are nonzero in the DFT Hamiltonian, the magnetic moments tilt toward the $a$ and $c$ axes even when a magnetic field is applied along the $b$ axis.
On the other hand, the magnetic moment along the $a$ axis has not been observed in experiments.
We tune the CEF parameters to eliminate the magnetic moment along the $a$ axis.
Specifically, we set the coefficients of the $O_{2,-2}\propto xy$, $O_{4,-4}\propto xy(x^{2}-y^{2})$, $O_{4,-2}\propto xy(7z^{2}-r^{2})$, $O_{2,1}\propto zx$, $O_{4,1}\propto zx(7z^{2}-3r^{2})$, and $O_{4,3}\propto zx(x^{2}-3y^{2})$ terms to zero.
Here, the $x$, $y$, and $z$ directions are aligned with the crystallographic $a$, $b$, and $c$ axes, respectively.

Next, we formulate strong $f$-electron correlations in the DMFT framework.
To describe the completely localized limit considered in this work, we adopt the Hubbard-I approximation.
For simplicity, we omit the $j=7/2$ states of $4f$ electrons, which lie $\Delta_{\mathrm{SOC}}\simeq 0.33$~eV above the $j=5/2$ states because of spin-orbit coupling.
The DFT+Hubbard-I calculations are applied to the remaining 360 orbitals.
The single-particle Green's function is defined as 
\begin{equation}
  \hat{G}(\bm{k},z)=[(z+\mu)\hat{1}-\hat{H}_{\mathrm{TB}}(\bm{k})-\hat{\Sigma}_{\mathrm{loc}}(z)-\hat{\Sigma}_{\mathrm{pot}}]^{-1}.
  \label{eq:Green_function_SM}
\end{equation}
$\hat{H}_{\mathrm{TB}}(\bm{k})$ is the tight-binding Hamiltonian with the tuned CEF parameters.
$\hat{\Sigma}_{\mathrm{loc}}(z)$ is the local self-energy of the $f$ electrons in the Hubbard-I approximation.
The last term, $\hat{\Sigma}_{\mathrm{pot}}$, in Eq.~\eqref{eq:Green_function_SM} is a local potential term for the $f$ electrons.
We set 
\begin{equation}
    \hat{\Sigma}_{\mathrm{pot}}=\Delta\epsilon_{f}\hat{\mathcal{P}}_{f}-h\hat{M}_{b}^{\mathrm{AFM}}-h_{c}\hat{M}_{c}^{\mathrm{FM}}.
    \label{eq:pot_SM}
\end{equation}
Here, $\Delta\epsilon_{f}$ is the energy shift of the $4f$-level, $\hat{\mathcal{P}}_{f}$ is the projection operator onto the $4f$-orbitals, $h$ is the molecular field strength, and $h_{c}$ is the uniform magnetic field strength along the $c$ axis.
The $\Delta\epsilon_{f}$ term in $\hat{\Sigma}_{\mathrm{pot}}$ accounts for double counting of the $4f$ electrons.
The second term is a molecular field term to induce the collinear AFM order along the $b$ axis, where $\hat{M}_{b}^{\mathrm{AFM}}$ is the corresponding configuration in Fig.~\ref{fig:electronic_struct_SM}(a).
The third term controls the canting magnetization along the $c$ axis, where $\hat{M}_{c}^{\mathrm{FM}}$ is the uniform ferromagnetic moment along the $c$ axis.

We compute the self-energy $\hat{\Sigma}_{\mathrm{loc}}(z)$ using the exact diagonalization method within the Hubbard-I approximation.
We adopt the fully rotationally invariant Coulomb interaction with the conventional parametrization~\cite{anisimovFirstprinciplesCalculationsElectronic1997}.
These calculations are performed using the open-source software \texttt{DCore}~\cite{shinaokaDCoreIntegratedDMFT2021}, implemented with \texttt{TRIQS}~\cite{parcolletTRIQSToolboxResearch2015} and \texttt{DFTTools}~\cite{aichhornTRIQSDFTToolsTRIQS2016}.
The impurity problem is solved by exact diagonalization using \texttt{pomerol}~\cite{andrey_e_antipov_2017_825870}.
The number of $\bm{k}$ points in the Hubbard-I-based calculations is set to $12\times12\times6$, whereas the number of positive Matsubara frequencies is set to $2048$.

Figure~\ref{fig:akw_SM} shows the single-particle spectral function $A(\bm{k},\omega)=-\frac{1}{\pi}\mathrm{Im}\mathrm{Tr}\hat{G}(\bm{k},\omega)$ in the paramagnetic state for \cecuge.
The excitation energy of the $4f$ electrons $4f^{n}\!\to\!4f^{n\pm1}$ with $n=1$ is denoted by $\Delta_{\pm}$.
We set $\Delta_{+}=0.31$~eV and $\Delta_{-}=0.19$~eV, which are the values for the cerium element~\cite{herbst$4f$ExcitationEnergies1976,herbstRelativisticCalculations$4f$1978}.
To reproduce these values, we set $\Delta\epsilon_{f}=-2.04$~eV in Eq.~\eqref{eq:pot_SM}, the direct on-site Coulomb interaction $U=6.17$~eV, and the Hund's coupling $J=0.79$~eV~\cite{lochtStandardModelRare2016}.

\begin{figure}[tb]
  \centering
  \includegraphics[width=0.8\linewidth]{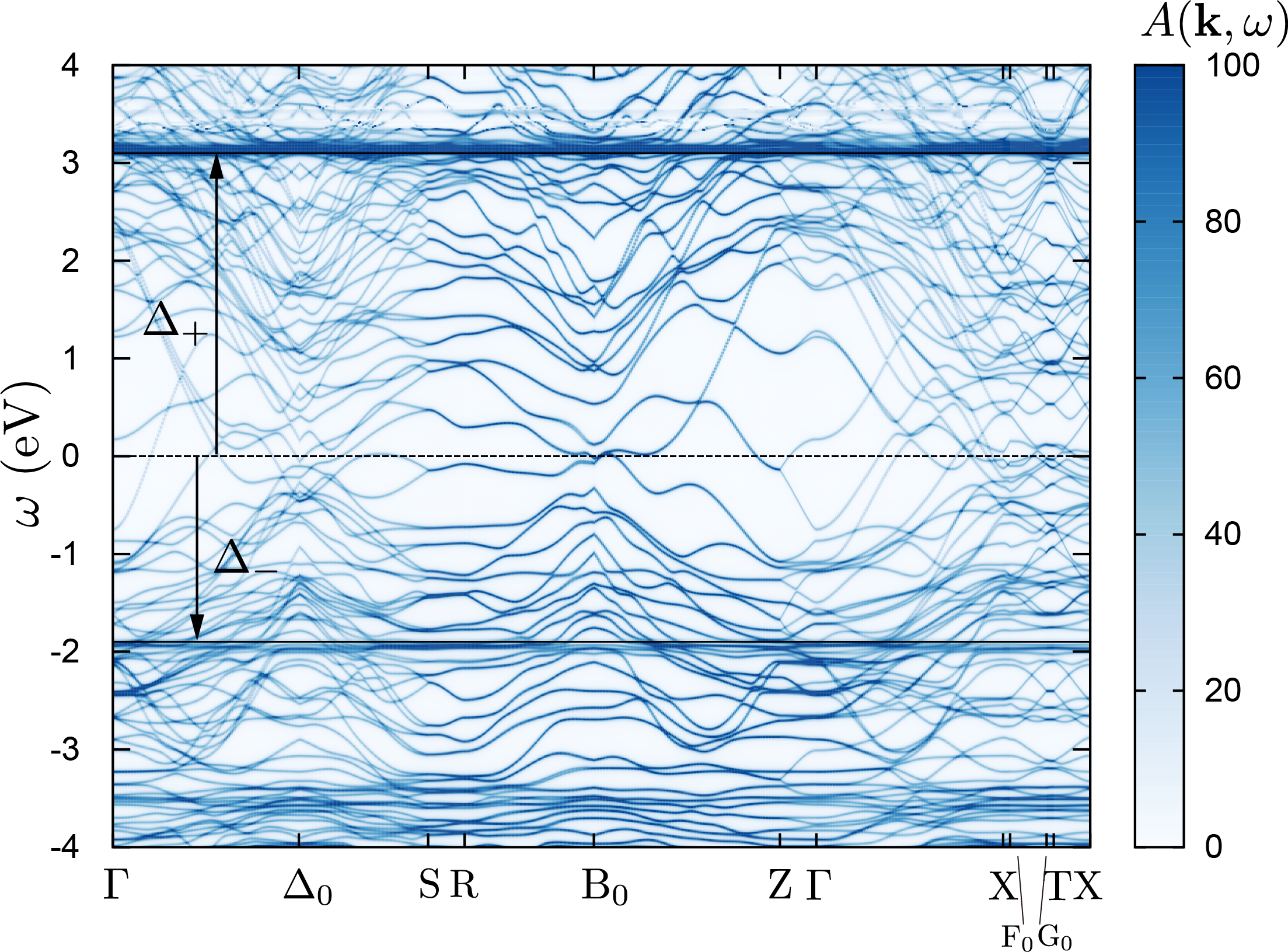}
  \caption{Single-particle spectral function in the paramagnetic state for \cecuge.
  $\Delta_{\pm}$ represents the excitation energy of the $4f$ electrons $4f^{n}\!\to\!4f^{n\pm1}$ with $n=1$.}
  \label{fig:akw_SM}
\end{figure}

\section{Transport properties}

\subsection{Green's-function evaluation of the intrinsic anomalous Hall conductivity}

By using the single-particle Green's function in Eq.~\eqref{eq:Green_function_SM}, we can evaluate the intrinsic AHC from the current--current correlation function~\cite{tanakaTheoryAcAnomalous2008,parkerDiagrammaticApproachNonlinear2019,michishitaEffectsRenormalizationNonHermiticity2021}.
The current--current correlation function within the local-self-energy framework is given by
\begin{equation}
  \Phi_{xy}(\tau)=\int\frac{d\bm{k}}{(2\pi)^{3}}\mathrm{Tr}[\hat{\mathcal{J}}_{x}(\bm{k})\hat{G}(\bm{k},\tau)\hat{\mathcal{J}}_{y}(\bm{k})\hat{G}(\bm{k},\beta-\tau)]
  \label{eq:current_current_correlation_function_SM}
\end{equation}
in the imaginary-time domain $\tau\in(0,\beta)$, corresponding to the bubble diagram.
$\hat{\mathcal{J}}_{\mu}(\bm{k})=(e/\hbar)(\partial/\partial k_{\mu})\hat{H}_{\mathrm{TB}}(\bm{k})$ for $\mu=x,y$ is the current operator.
$e$ and $\hbar$ are the elementary charge and reduced Planck constant, respectively.
$\mathrm{Tr}$ includes the trace over the site, orbital, and spin indices within a unit cell.
The DC conductivity is given by
\begin{equation}
  \sigma_{xy}^{\mathrm{DC}}=\lim_{\omega\to 0}\frac{\Phi_{xy}^{\mathrm{R}}(\omega)-\Phi_{xy}^{\mathrm{R}}(0)}{i\omega},
  \label{eq:DC_conductivity_SM}
\end{equation}
where $\Phi_{xy}^{\mathrm{R}}(\omega)$ is the retarded current--current correlation function in the real-frequency domain.
We use the intermediate representation (IR) basis~\cite{shinaokaCompressingGreensFunction2017} to perform the Fourier transformation of the Green's function between the Matsubara-frequency and imaginary-time domains in Eq.~\eqref{eq:Green_function_SM} [Eq.~\eqref{eq:current_current_correlation_function_SM}].
The IR basis set is generated by the \texttt{sparse-ir} library~\cite{wallerbergerSparseirOptimalCompression2023} based on the sparse sampling~\cite{liSparseSamplingApproach2020}.
We obtain the DC conductivity by (i) performing analytic continuation of the current--current correlation function from Matsubara to real frequency using the Pad\'{e} approximant method and (ii) fitting the optical Hall conductivity with the Drude model in the low-frequency region:
\begin{equation}
  \sigma_{xy}(\omega)=\frac{\sigma_{xy}^{\mathrm{DC}}}{(1-i\omega\tau_{0})^{2}},
\end{equation}
where $\tau_{0}$ stands for the relaxation time.
The integral is performed over $192\times 192\times 128$ $\bm{k}$-points in Eq.~\eqref{eq:current_current_correlation_function_SM}.
We set $h=\pm 0.1$~eV and vary $h_{c}$ from $0$ to $2.0$~meV in Eq.~\eqref{eq:pot_SM} to study how canting magnetization affects the intrinsic AHC.

\subsection{Distribution of the Berry curvature}

We obtain the intrinsic AHC from the momentum-space integral of the Berry curvature.
The band-resolved Berry curvature $\Omega_{n\bm{k}}^{z}$ is defined as~\cite{xiaoBerryPhaseEffects2010}
\begin{equation}
 \Omega_{n\bm{k}}^{z}=-2\hbar^{2}\sum_{m(\neq n)}\frac{\mathrm{Im}[v_{x\bm{k}}^{nm}v_{y\bm{k}}^{mn}]}{(\epsilon_{n\bm{k}}-\epsilon_{m\bm{k}})^{2}}.
 \label{eq:Berry_curvature_SM}
\end{equation}
Here, $\epsilon_{n\bm{k}}$ is the eigenvalue of the effective Hamiltonian for the $n$-th band at $\bm{k}$.
$v_{x\bm{k}}^{nm}$ and $v_{y\bm{k}}^{nm}$ are the Bloch representations of the velocity operators $\hat{v}_{\mu\bm{k}}=\hat{\mathcal{J}}_{\mu}(\bm{k})/e$ for $\mu=x,y$.
Figure~\ref{fig:Berry_curvature_SM} shows the distribution of the Berry curvature $\Omega_{\bm{k}}^{z}=\sum_{n}f_{n\bm{k}}\Omega_{n\bm{k}}^{z}$ in the Brillouin zone for the effective Hamiltonian.
Its distribution is localized near the spin-split band.
Figure~\ref{fig:Berry_curvature_SM}(b) shows that the Berry-curvature distribution from \ce{Ge}~$4p$ orbitals is consistent with that of the full effective Hamiltonian in Fig.~\ref{fig:Berry_curvature_SM}(a).
When we separate the contributions into the local effective field and nonlocal spin-dependent hopping, the local-field term is negligible and the hopping term dominates [Figs.~\ref{fig:Berry_curvature_SM}(c) and \ref{fig:Berry_curvature_SM}(d)].
Thus, the intrinsic AHC in the localized AFM structure arises primarily from nonlocal spin-dependent hopping.

\begin{figure}[tb]
  \centering
  \includegraphics[width=0.8\linewidth]{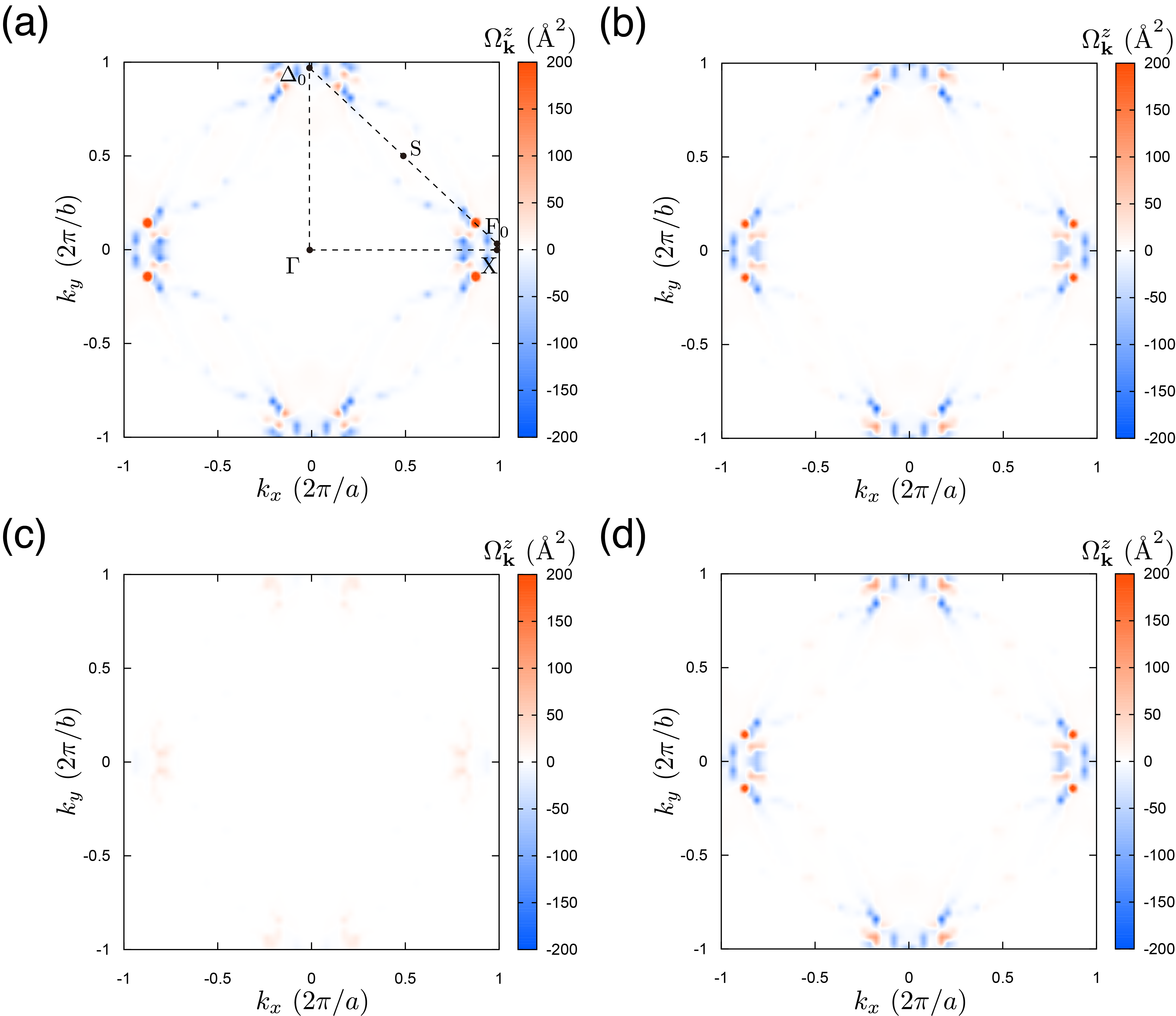}
  \caption{Berry-curvature distribution in the Brillouin zone at $k_{z}=0$.
  (a) Full effective Hamiltonian.
  (b) \ce{Ge}~$4p$ contribution.
  (c) Local-effective-field contribution from \ce{Ge}~$4p$ orbitals.
  (d) Nonlocal-spin-dependent-hopping contribution from \ce{Ge}~$4p$ orbitals.}
  \label{fig:Berry_curvature_SM}
\end{figure}

\subsection{Hopping-path dependence of the intrinsic AHC}

We determine which Ge-hopping paths contribute most to the intrinsic AHC.
As shown in Table~\ref{tab:wyckoff_positions_SM}, there are five inequivalent \ce{Ge} sites in the unit cell.
Among them, the \ce{Ge}~1 and \ce{Ge}~2 atoms form a zigzag chain, whereas the \ce{Ge}~3, \ce{Ge}~4, and \ce{Ge}~5 atoms form a two-dimensional bilayer structure.
Figure~\ref{fig:LC_band_path_SM} shows the intrinsic AHC obtained by including the $c$--$f$ hybridization effect of the \ce{Ge}~$4p$ orbitals in the zigzag chain and bilayer structure independently.
Near the Fermi level, the bilayer structure shows a larger AHC than the zigzag chain.
On the other hand, even if we sum up the contributions of the zigzag chain and bilayer structure, the AHC for the \ce{Ge}~$4p$ orbitals, considering all paths, is not fully reproduced.
This indicates that the hopping between the zigzag chain and bilayer structure is also important.
Therefore, the intrinsic AHC in this system is attributed to the three-dimensional network structure of the \ce{Ge} sites.
These results support the importance of the nonlocal spin-dependent hopping derived from the $c$--$f$ virtual process.

\begin{figure}[tb]
  \centering
  \includegraphics[width=\linewidth]{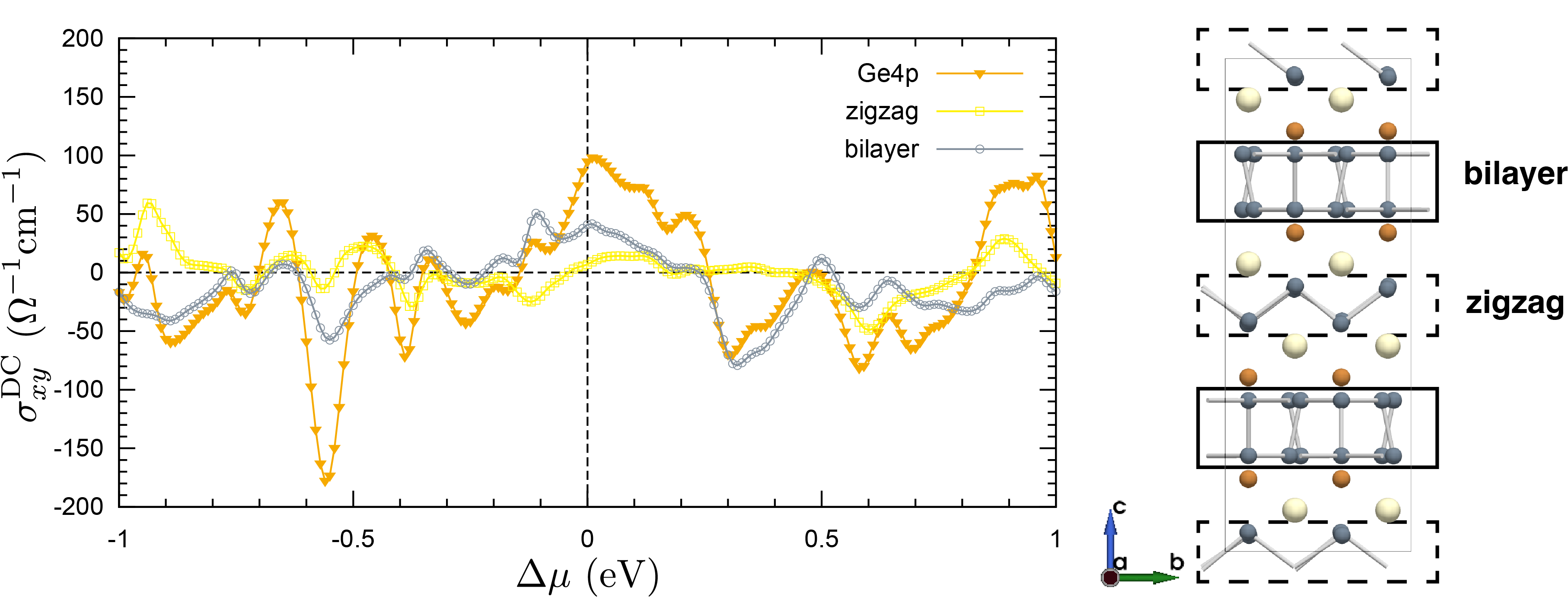}
  \caption{Hopping-path dependence of the intrinsic AHC from \ce{Ge}~$4p$ orbitals.
  Dashed lines indicate zigzag chains formed by \ce{Ge}~1 and \ce{Ge}~2.
  Solid lines indicate the bilayer structure formed by \ce{Ge}~3, \ce{Ge}~4, and \ce{Ge}~5 (See Table~\ref{tab:wyckoff_positions_SM}).}
  \label{fig:LC_band_path_SM}
\end{figure}

\end{widetext}

\end{document}